\documentclass[10pt,pra,nofootinbib,notitlepage,twocolumn,superscriptaddress]{revtex4-1}

\usepackage[utf8]{inputenc}
\usepackage[T1]{fontenc}
\usepackage[english]{babel}  
\usepackage{float}
\usepackage{times}
\usepackage{dsfont}
\usepackage{hyperref,amsmath,amssymb,amscd}
\usepackage{amssymb,amsmath,amsfonts,amsthm}
\usepackage{mathtools}
\usepackage{verbatim}
\usepackage{enumerate}
\usepackage{graphicx}
\graphicspath{{figs/}}
\usepackage{hyperref}
\usepackage{bbm}
\usepackage{booktabs}

\usepackage[caption=false]{subfig}

\usepackage{color}
\definecolor{dred}{rgb}{.8,0.2,.2}
\definecolor{ddred}{rgb}{.8,0.5,.5}
\definecolor{dblue}{rgb}{.2,0.2,.8}
\definecolor{dgreen}{rgb}{.2,0.5,.2}

\newcommand{\bra}[1]{\mbox{$\langle #1|$}}
\newcommand{\ket}[1]{\ensuremath{|#1\rangle}}

\newcommand{\be}{\begin{equation}}
\newcommand{\ee}{\end{equation}}

\newcommand{\bea}{\begin{eqnarray}}
\newcommand{\eea}{\end{eqnarray}}

\begin{document}

\title{Implementation of Multiparty quantum clock synchronization}

\author{Xiangyu Kong}
\affiliation{State Key Laboratory of Low-Dimensional Quantum Physics and Department of Physics, Tsinghua University, Beijing 100084, China}

\author{Tao Xin}
\affiliation{State Key Laboratory of Low-Dimensional Quantum Physics and Department of Physics, Tsinghua University, Beijing 100084, China}

\author{ShiJie Wei}

\affiliation{State Key Laboratory of Low-Dimensional Quantum Physics and Department of Physics, Tsinghua University, Beijing 100084, China}

\author{Bixue Wang}

\affiliation{State Key Laboratory of Low-Dimensional Quantum Physics and Department of Physics, Tsinghua University, Beijing 100084, China}

\author{Yunzhao Wang}

\affiliation{State Key Laboratory of Low-Dimensional Quantum Physics and Department of Physics, Tsinghua University, Beijing 100084, China}
\author{Keren Li}

\affiliation{State Key Laboratory of Low-Dimensional Quantum Physics and Department of Physics, Tsinghua University, Beijing 100084, China}

\author{GuiLu Long}
\affiliation{State Key Laboratory of Low-Dimensional Quantum Physics and Department of Physics, Tsinghua University, Beijing 100084, China}

\affiliation{
The Innovative Center of Quantum Matter, Beijing 100084, China}

\begin{abstract}

The quantum clock synchronization (QCS) is to measure the time difference among the spatially separated clocks with the principle of quantum mechanics. The first QCS algorithm proposed by Chuang and Jozsa is merely based on two parties, which is further extended and generalized to the multiparty situation by Krco and Paul. They present a multiparty QCS protocol based upon W states that utilizes shared prior entanglement and broadcast of classical information to synchronize spatially separated clocks. Shortly afterwards, Ben-Av and Exman came up with an optimized multiparty QCS using Z state. In this work, we firstly report an implementation of Krco's and Ben-AV's multiparty QCS algorithm using a four-qubit Nuclear Magnetic Resonance (NMR). The experimental results show a great agreement with the theory and also prove Ben-AV’s multiparty QCS algorithm
more accurate than Krco's.
\end{abstract}

\maketitle

\section{Introduction}
The clock synchronization, which determines the time difference among the spatially separated clocks, is an important issue with many practical and scientific applications~\cite{ap1,ap2}.  As the heart of many modern technologies, clock synchronization is instrumental to those area such as  global positioning system (GPS), electric power generation (synchronization of generators feeding into national power grids) and telecommunication (synchronization data transfers, financial transactions). Scientifically, clock synchronization is the key to projects such as long baseline interferometry (distributed radio telescopes), gravitation wave observation (LIGO), tests of general theory of relativity, and distributed computation.

 Current approaches to clock synchronization are mainly based on two classical protocols which are proposed by Einstein~\cite{Einstein} and Eddington~\cite{Addington}. Assuming that there are two spatially separated clocks A and B which are at rest in a common inertial frame in the special theory of relativity. Einstein Synchronization involves an operational line-of-sight exchange of light pulses between two observers Alice and Bob who are collocated with their clocks A and B separately~\cite{Anderson}. As for the other method, Eddington's slow clock transport demands that the two clocks A and B should be first synchronized locally and then are transported adiabatically to their final separate location. However, in these clock synchronization protocols, actual timing information must be transferred from one clock to another over some channels, whose imperfections generally limit the accuracy of the synchronization~\cite{Jozsa}.

As quantum information flourishes these years, the quantum entanglement is considered as a precious resource extensively investigated for a variety of applications in distributed systems, for instance, quantum key distribution (QKD)~\cite{QKD} and quantum secure direct communication (QSDC)~\cite{QSDC1,QSDC2,QSDC3}. Moreover, it would be more fun to use a method that exploits the resource of quantum
entanglement, what the JPL group called Quantum
(Atomic) Clock Synchronization (QuACS or QCS) ~\cite{preskill}. Jozsa \textit{et al} first purposed the basic protocol for the synchronization of two spatially separated parties based upon shared prior quantum entanglement. For its importance, Valencia \textit{et al }~\cite{Valencia}  and Quan \textit{et al}~\cite{Quan} reported their two-party clock synchronization experiments with different distance in succession. Another two-party QCS algorithm proposed by Chuang ~\cite{chuang} had been implemented in NMR system by Zhang \textit{et al} ~\cite{Long}. Based on the two-party ideas, general frameworks for multiparty clock synchronization were further proposed by Krco \textit{et al} ~\cite{Krco} , Ben-Av \textit{et al} ~\cite{Ben} and Ren \textit{et al} ~\cite{Ren} respectively. We focus on the first two protocols. The two protocols begin with different initial states: one is W state and the other depends on the size of system. We will introduce these two QCS protocols in detail in next section. 

In this work, for the first time, we experimentally demonstrate the multiparty quantum clock synchronization protocols in a four-qubit NMR system, where one qubit is used as referential clock, and the other three qubits act as the clocks to be synchronized.

The paper is organized as follows: In Sec. \ref{theory}, we briefly review the basic quantum clock synchronization and the multiparty QCS protocols. In Sec. \ref{experiment and result},  we introduce our experimental setups and experimental procedure.  Then, we present the experimental results and discuss the consequences. Finally, Sec. \ref{conclusion} summarizes the entire work and give some prospects in its future applications.

\section{theoretical overview}
\label{theory}

 \textit{Basic quantum clock synchronization}--- Consider a qubit with stationary states \ket{0} and \ket{1} having the energy eigenvalues $E_0 < E_1$, respectively. Then we introduce the dual basis $\ket{pos}=(1/\sqrt{2})(\ket{0}+\ket{1})$ and $\ket{neg}=(1/\sqrt{2})((\ket{0}-\ket{1})$ with the angular frequency $\omega =\frac{1}{\hbar}(E_1-E_0)$.  At the beginning of the protocol, we prepare the initial entangled state as
   \begin{equation}
   \label{eq:initial state}
   \ket{\psi}=\frac{\ket{00}+\ket{11}}{\sqrt{2}}=\frac{\ket{pos}_A\ket{pos}_B+\ket{neg}_A\ket{neg}_B}{\sqrt{2}}.
   \end{equation}
where A and B stand for the subsystem distributed to Alice and Bob. We assumed that the clock of Alice is the standard one. The state in the hand of Bob would immediately collapse to $\ket{pos}$  if the measured result on Alice's particle is $\ket{pos}$ at $t_A=0$. However, since the time difference $\Delta$ between Alice and Bob, when Alice measures the particle at hand with $t_A=0$, the clock of Bob is not zero with $t_B\neq0$. If the clock of Bob is behind, the state of Bob when $t_B=0$ is
   \begin{equation}
   \label{eq:ending state}
   \ket{\psi_B}=\frac{e^{i\omega\Delta/2}\ket{0}+e^{-i\omega\Delta/2}\ket{1}}{\sqrt{2}}.
   \end{equation}
According to the theoretical protocol, Bob will measure his particle  in the measurement basis at time $t_B=0$ as usual.  If the result of  Alice's measurement is $\ket{pos}$, Bob will obtain the possible outcome with the following probabilities:
   \begin{align}
   \label{eq:probability}
   P(pos)=\langle {pos}\ket{\psi_B}=\frac{1+\cos(\omega\Delta)}{2}.
   \end{align}
It is concluded that the relative probability of the result (\ket{pos} can be used to estimate $\Delta$ with the condition $|\omega\Delta|<2\pi$. Hence, a general framework of multiparty clock synchronization protocols is constructed based on the above idea.

\begin{figure}[h]
\includegraphics[scale=0.4]{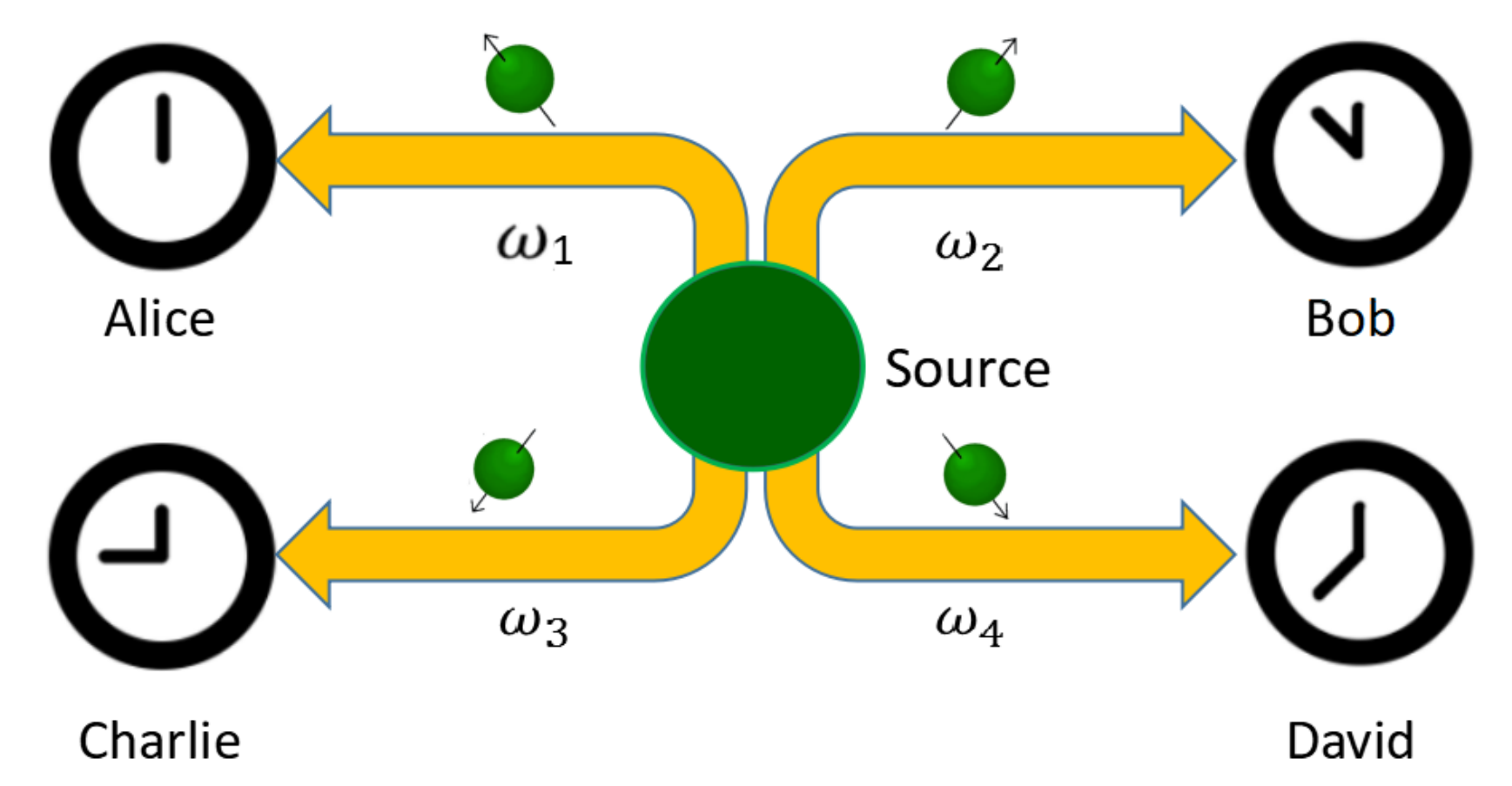}
\caption{The schematic of four parties QCS protocol. Alice, Bob, Charlie and David with their own clocks locate separately. The green area in the middle is a source library which distributes the entangled particles, where $\omega_{1}$, $\omega_{2}$, $\omega_{3}$ and $\omega_{4}$ represent the frequency of different particles. }
\label{fig:4QCS}
\end{figure}

 \textit{Two multiparty QCS protocols}---Multiparty protocols aim at synchronizing $n$ spatially separated clocks, any one of which can be later taken as the standard clock. The basic idea is illustrated with a specific case where $n=4$ in Fig. \ref{fig:4QCS}. Four clocks without synchronization in spatially separated areas have their own observer respectively: Alice, Bob, Charlie and David. Since the time difference between any two clocks, In general, we take the clock of Alice as the standard clock. In the middle green area, there is a particle source library which provides four-qubit entangled particles. These entangled particles with different spin directions are passed to the four observers. Through their measurement on the particles at the same time as shown by their own clock. These four parties can coordinate their clocks according to the measuring result based on above idea. The most popular multiparty synchronization protocol are presented by Krco and Ben-Av, and will be illustrated in the following.

 \textbf{Protocol A}: The protocol was present by Krco starts from an initial W state (see,e.g., D\"{u}r \textit{at al } ~\cite{wstate}). W states are entangled states which have $n$ terms in following form:
 \begin{align}
   \label{eq:w state}
  \ket{\psi(W)}=\frac{1}{\sqrt{n}}\underbrace {\left( \ket{10...00}+\ket{01...00}+...+\ket{00...01} \right)}_{n \ terms}
\end{align}
where each term contains only a single qubit in the state $\ket{1}$. Without loss of the generality, we take the clock of Alice as the standard one and synchronize the others. It is similar  with the two-party protocol process. Firstly, at standard time $t_A=0$, Alice measures the qubit in her possession in the measurement basis ($\ket{pos},\ket{neg}$). Secondly, she will announce classically the measurement results. The observers of the other clocks without synchronizing , also measure their own qubits in the measurement basis at their own time $t=0$, which has a time difference  $\Delta$ from the standard time.

For the case that the result measured by Alice is $\ket{pos}$, the others obtain the probabilities $P$ with the different outcomes:
 \begin{align}
   \label{eq:w_result}
   P(\ket{pos})=\frac{1}{2}+\frac{\cos(\omega\Delta)}{n},\omega=\omega_2,\omega _3,...,\omega_n.
\end{align}
Hence the relative probabilities of the results will enable the others to estimate $\Delta$ and adjust their clocks assuming that $|\omega\Delta|<2\pi$.

Krco and Paul point out that the accuracy of determination of $\Delta$ decreases with $n$ because the amplitude of the probability decreases with $n$ in the Eq. (\ref{eq:w_result}). They attribute it to the fact that the entanglement of the initial state decrease with $n$, in Eq. (\ref{eq:w_result}). Therefore Ben-Av \textit{et al} introduce a better initial state where the accuracy of determination of $\Delta$ decreases with $n$ more slowly and the limit of the amplitude is not zero when $n$ approaches infinity.

\textbf{Protocol B}: Ben-Av extended the protocol by providing a different initial state. This initial state is denoted by $\psi(Z_n^k)$, where $n$ is the total number of qubits (particles) and $k$ is the number of qubits in the $\ket{1}$ state in each term, as follows:
 \begin{gather}
   \ket{\psi(Z_n^k)}=A\underbrace{\left( \ket{11...00}+\ket{10...1...0}+...+\ket{00...11}\right)}_{n\ terms}\notag\\
  A={\left(\sqrt{\frac{n!}{(n-k)!k!}}\right)}^{-1}.
  \label{eq:z state}
\end{gather}

In the similar with protocol A, we take the clock of Alice as the standard clock and synchronize the other clocks. Following the same process, at last for the sets measured by Alice as $\ket{pos}$, the others get the probabilities $P$ with the possible outcomes:
 \begin{gather}
   \label{eq:z_result}
P(\ket{pos})=\frac{1}{2}+\frac{k(n-k)}{n(n-1)}\cos(\omega\Delta),\omega=\omega_2,\omega _3,...,\omega_n,\notag\\
A_0(n,k)=\frac{k(n-k)}{n(n-1)}.
\end{gather}
\begin{figure*}[htbp]
\begin{center}
\includegraphics[width= 2\columnwidth]{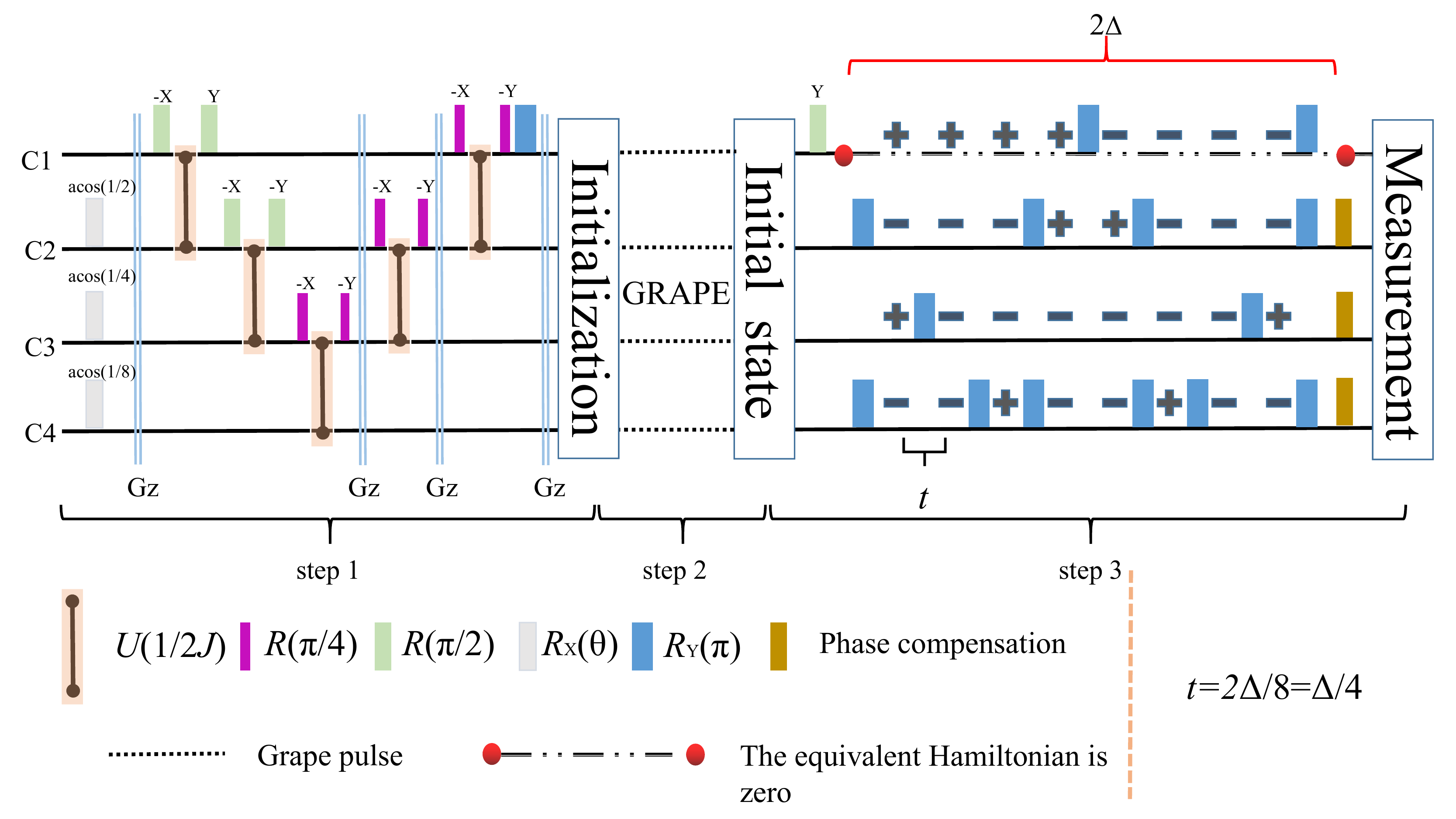}
\end{center}
\setlength{\abovecaptionskip}{-0.00cm}
\caption{\footnotesize{\textbf{NMR sequence to realize the four parties QCS protocols.} In the first step, we use spatial averaging to prepare the pseudo-pure state. In this part, the purple rectangles stand for the unitary operations which rotate the forty-five degrees with a coordinate axis. Similarly, the green rectangles rotate ninety degrees. The white rectangles stand for the unitary operations which rotate the special angle with X axis. The blue rectangles stand for the unitary operations which rotate one hundred and eighty degrees with Y axis. The double straight lines stand for the gradient field in the z direction. In the second step, the radio-frequency (RF) pulses during this procedure are optimized by the GRAPE technology to create the initial states of the different protocols. In the third step, this sequence is used to change the experiment Hamiltonian as showed in Eq. (\ref{Hamiltonian ideal}). In this part, each positive sign or negative sign stands for the equal $t$ of the evolution time: $t=\Delta/4$ which is proved in Eq. (\ref{eq:echo}).}} \label{fig:quantum circuit}
\end{figure*}

To improve the clock adjustment accuracy we can choose an optimal $k$ for a given $n$ as follows. We denote $A_0$ as the amplitude of the time probability fluctuation. Obviously, for any given $n$ we wish to choose $k$ such that $A_0$ is maximized. One can easily see that $k_{opt}$ \ (optimal $k$) is
 \begin{gather}
   \label{eq:optimal n}
   k_{opt}=\lfloor{n/2}\rfloor\\
    A_0=\frac{\lfloor{n/2}\rfloor*\lfloor{n/2}\rfloor}{n(n-n)}
\end{gather}

For $n\geqslant4$ this result is a clear improvement over the original W state ($k=1$), for which $A_0(1,n)=1/n$.

The outline of our approach is as follows. Just as we have introduced before, we focus on the four-qubit protocol. Thus the probabilities of the two protocols can be assigned as:
\begin{align}
   \label{eq:4bit-wstate-result}
   P_A(\ket{pos})=\frac{1}{2}+\frac{\cos(\omega\Delta)}{4},\omega=\omega_2,\omega _3,\omega_4,\\
   \label{eq:4bit-zstate-result}
   P_B(\ket{pos})=\frac{1}{2}+\frac{\cos(\omega\Delta)}{3},\omega=\omega_2,\omega _3,\omega_4.
\end{align}
We take the clock of Alice as the standard clock, and set various time difference between Alice and the others. Then we will check whether the probabilities we get from experiments are agreed with the corresponding theory equations. Moreover, we will also prove whether the accuracy of the protocol B is better than protocol A.

\section{experiment and result}
\label{experiment and result}
We experimentally inspect the two multiparty protocols using nuclear magnetic resonance (NMR).
The four-qubit sample is $^{13}$C-labeled trans-crotonic acid dissolved in d6-acetone.
\begin{figure}[htb]
\begin{center}
\includegraphics[width= 1\columnwidth]{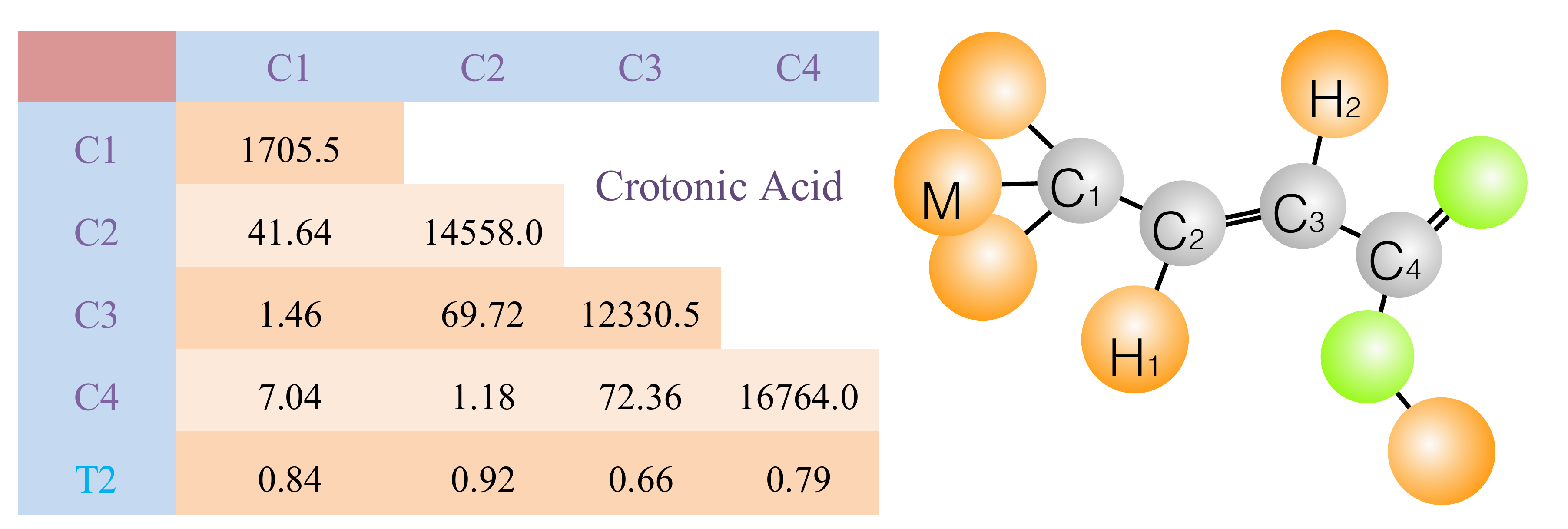}
\end{center}
\setlength{\abovecaptionskip}{-0.00cm}
\caption{\footnotesize{  Molecular structure and Hamiltonian parameters of $^{13}$C-labeled trans-crotonic acid. C$_1$, C$_2$, C$_3$ and C$_4$ are used as four qubits. The chemical shifts and J-couplings (in Hz) are listed by the diagonal and off-diagonal elements, respectively.  T$_{2}$ (in Seconds) are also shown at bottom.}}\label{fig:molecule}
\end{figure}
The structure of the molecule is shown in Fig. \ref{fig:molecule}, where C$_1$ to C$_4$ denote the four qubits, representing the clocks of Alice, Bob, Charlie and David in protocol. The methyl group M, H$_1$ and H$_2$ were decoupled throughout all experiments. The internal Hamiltonian under weak coupling approximation is
\begin{align}\label{Hamiltonian}
\mathcal{H}=-\sum\limits_{j=1}^4 {\frac{1}{2} \omega _j } \sigma_z^j  + \sum\limits_{j < k}^4 {\frac{\pi}{2}} J_{jk} \sigma_z^j \sigma_z^k,
\end{align}
where $\nu_j$ is the chemical shift and $\emph{J}_{jk}$ is the J-coupling strength. All experiments were carried out on a Bruker DRX 400MHz spectrometer at room temperature (296.5K).

The entire experiment can be divided into three parts and its experimental circuits are shown in Fig. \ref{fig:quantum circuit}.

\begin{figure}[htb]
\begin{center}
\includegraphics[width= 0.95\columnwidth]{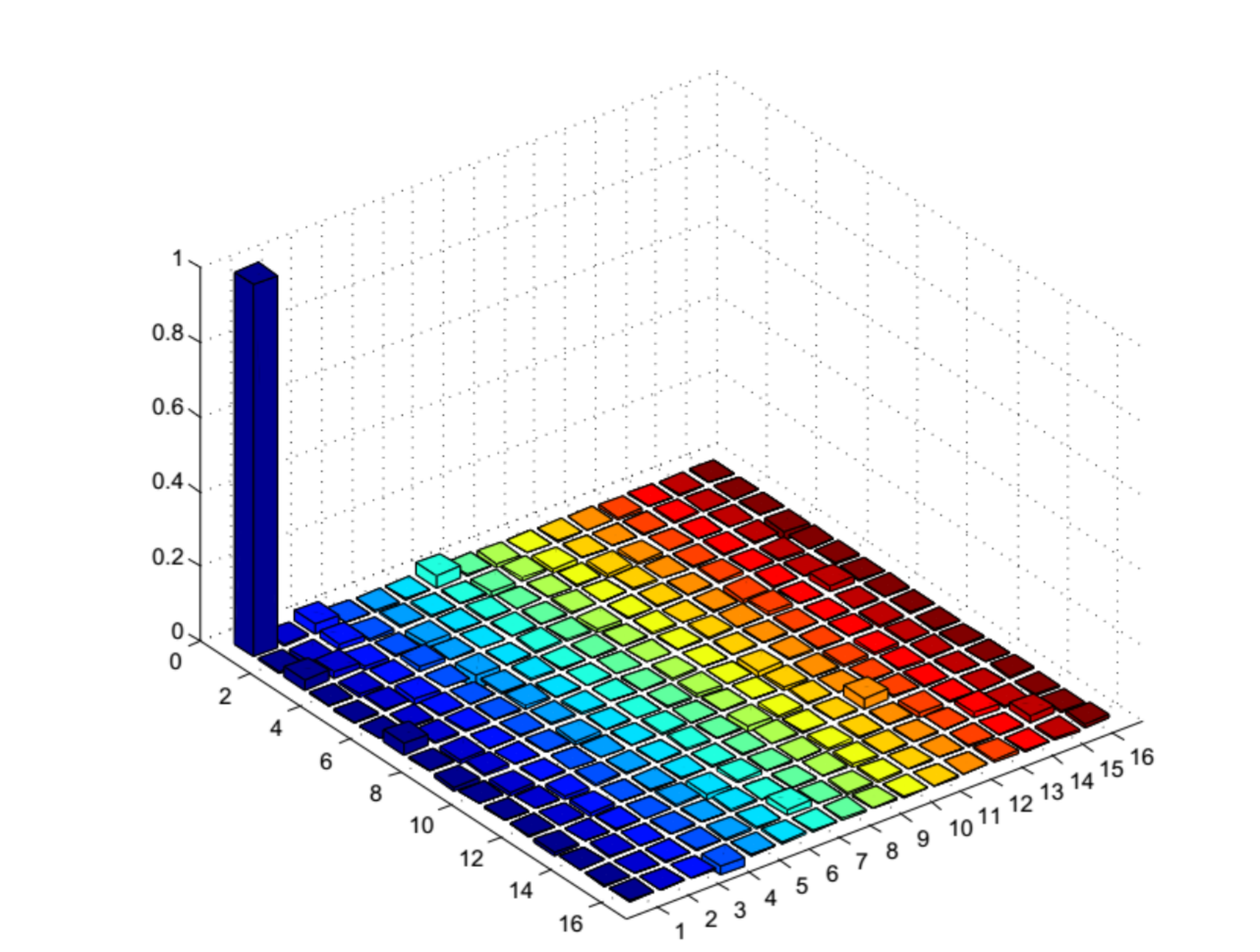}
\end{center}
\setlength{\abovecaptionskip}{-0.00cm}
\caption{\footnotesize{\textbf{The density matrix of the experimental results of \ket{0000} (in the normalization units).} It is reconstructed using the state tomography technique. And only the real components are plotted. The imaginary portions from the experimental results, which are theoretically zero, are found to contribute less than 3.4\%. }} \label{fig:pps}
\end{figure}

\textit{Step1:Initialization}---Starting from thermal equilibrium state, we drive the system to the pseudo-pure state (PPS) with the method of the spatial averaging technique~\cite{cory,D. Lu,Y. Lu,Tao X,keren,Pearson}. Step $1$ in Fig. \ref{fig:quantum circuit} is the experimental circuit realizaing PPS where all local operations are optimized using the gradient ascent pulse engineering(Grape) with a fidelity over 99.5\%. The final form of four-qubit PPS is
$\rho_{0000}=(1-\epsilon){\mathbb{I}}/16+\epsilon\ket{0000}\bra{0000}$,
where $\mathbb{I}$ is identity and $\epsilon\approx 10^{-5}$ is the polarization. Since only the deviated part $\ket{0000}$ contributes to the NMR signals, the density matrix used in NMR are all deviated matrix and the PPS is able to serve as an initial state. The experimental results are represented as the density matrices obtained by the state tomography technique~\cite{tomo1,tomo2,tomo3} shown in Fig. \ref{fig:pps}. The fidelity between the experimental results with $\ket{0000}$  is over 99.02\%.

\textit{Step2:To the aim states}---Since the theoretical requirement of four entangled particles for each clock keeper, we evolve $\ket{0000}$ to the different entanglement state: W state or Z state depending on protocol A or B.

Protocol A: The initial state of this protocol is a four qubits W state, shown as:
\bea\label{eq:4bit-wstate}
\ket{\psi(W)}=\frac{1}{\sqrt{4}}(\ket{1000}+\ket{0100}+\ket{0010}+\ket{0001})
\eea

Protocol B: In this protocol, the initial entangled state is the optimal Z state. That is in the Eq. (\ref{eq:z state}), set $n=4$, $k=2$, shown as:
\bea\label{eq:4bit-zstate}
\ket{\psi(Z)}=\frac{1}{\sqrt{6}}(\ket{1100}+\ket{0110}+\ket{0011}+\notag\\
\ket{1010}+\ket{0101}+\ket{1001})
\eea
The experimental circuit is listed in Step 2 of Fig. \ref{fig:quantum circuit}. The radio-frequency (RF) pulses during this procedure are also optimized by the GRAPE technology~\cite{GRAPE1,GRAPE2}, and are designed to be robust to the static field distributions ($T_2^{*}$ process) and RF inhomogeneity. Moreover the designed fidelity for each pulse exceeds 0.995 with a length of 20ms.

\textit{Step3:Mesurement}---When the time of standard clock arrive at 0, we measure the qubit which stands for the standard clock respectively with basis ($\ket{pos}$,$\ket{neg}$). The measurement circuit is also listed in Step 3 at Fig. \ref{fig:quantum circuit}. In this step, the ideal Hamiltonian is depicted by Eq. (\ref{Hamiltonian ideal}). Without loss of generality, we assume that the clock of Alice is standard, and the other clocks are all slower. We have to set $\omega_{1}=0$ for the quantum state which we have measured do not change.
\begin{align}\label{Hamiltonian ideal}
\mathcal{H}_{\text{ideal}}=\sum\limits_{j=2}^4 {\frac{1}{2} \omega _j } \sigma_z^j \end{align}

Under such a Hamiltonian, we simulated the quantum state evolution with the $\Delta$ which stands for the time difference. Just as introduced before, the quantum state evolution is $\ket{\psi}(t)=e^{-i\omega t/2}\ket{\psi}(0)$.
Compared to the internal Hamiltonian in NMR system, the ideal Hamiltonian has no J-coupling term. This can be effectively simulated using the spin-echo technology ~\cite{echo}. Through a twice $\Delta$ length circuit shown in Step 3 of Fig. \ref{fig:quantum circuit}, we can get a effective Hamiltonian evolution without J-couplings and $\omega_{1}$. The details of calculation are shown in Eq. (\ref{eq:echo}).

\begin{align}\label{eq:echo}
&\quad \quad R^2_y(\pi)R^4_y(\pi)e^{-iH\Delta/4}R^3_y(\pi)e^{-iH
\Delta/4}R^4_y(\pi)e^{-iH\Delta/4}\notag\\&R^2_y(\pi)
R^4_y(\pi)e^{-iH\Delta/4}R^1_y(\pi)e^{-iH\Delta/4}
R^2_y(\pi)R^4_y(\pi)e^{-iH\Delta/4}\notag\\&R^4_y(\pi)
e^{-iH\Delta/4}R^3_y(\pi)R^1_y(\pi)e^{-iH\Delta/4}
R^2_y(\pi)R^4_y(\pi)\notag\\&
=e^{-iH_{\text{ideal}}\Delta}
\end{align}
At the end of circuit, we measure the probability for each clock of Bob, Charlie or David  with $\ket{pos}$ and $\ket{neg}$ basis. We take Bob as an example to introduce in detail. The frequency is $\omega_2/2\pi=250Hz$. Though $\omega_2/2\pi$ is quite different from the one given by the Fig. \ref{fig:molecule}, It can be corrected by a  phase compensation in the end of circuits shown in Fig. \ref{fig:quantum circuit}.

\begin{figure}[htb]
\begin{center}
\includegraphics[width= 1.05\columnwidth]{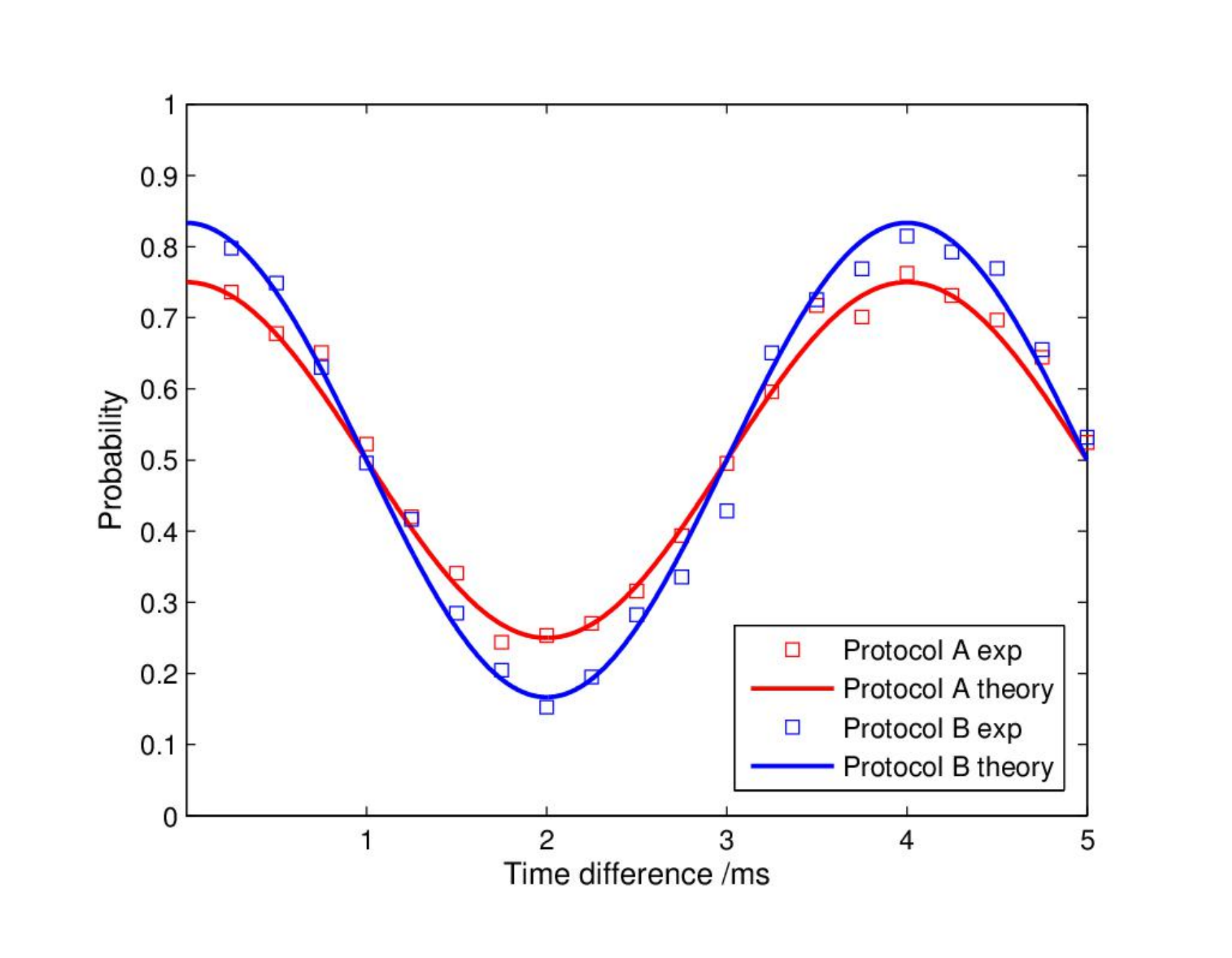}
\end{center}
\setlength{\abovecaptionskip}{-0.00cm}
\caption{\footnotesize{\textbf{Synchronize Bob's clock using different protocol.} The red line is the theory value of protocol A showed in Eq. (\ref{eq:4bit-wstate-result}) with $\omega_2/2\pi=250$, and the red squares are the experimental results of protocol A with 20 time difference $\Delta$. The blue line is the theory value of protocol B showed in Eq. (\ref{eq:4bit-zstate-result}) with $\omega_2/2\pi=250$, and the blue squares are the experimental results of protocol B with 20 time difference $\Delta$. }} \label{fig:result1}
\end{figure}

From 0 and 5 milliseconds, we choose 20 points at equal intervals as the time difference $\Delta$. Then we set the time difference $\Delta$ of Bob's clock and complete these experiments. Finally, we will get the probabilities in the experiment using the two protocols, and compare these points with the theory. The results are shown in Fig. \ref{fig:result1}.
The two different color squares which stand for the experimental results of the two protocols are in good agreement with the corresponding curves, whose functions are as shown in Eq. (\ref{eq:4bit-wstate-result}-\ref{eq:4bit-zstate-result}) with $\omega_2/2\pi=250Hz$. So the experiment is an excellent demonstration of the feasibility of the multiparty QCS. Moreover, It can be concluded that the results of protocol B is better than protocol A because of the larger amplitude.

\begin{figure}[htb]
\begin{center}
\includegraphics[width= 1.05\columnwidth]{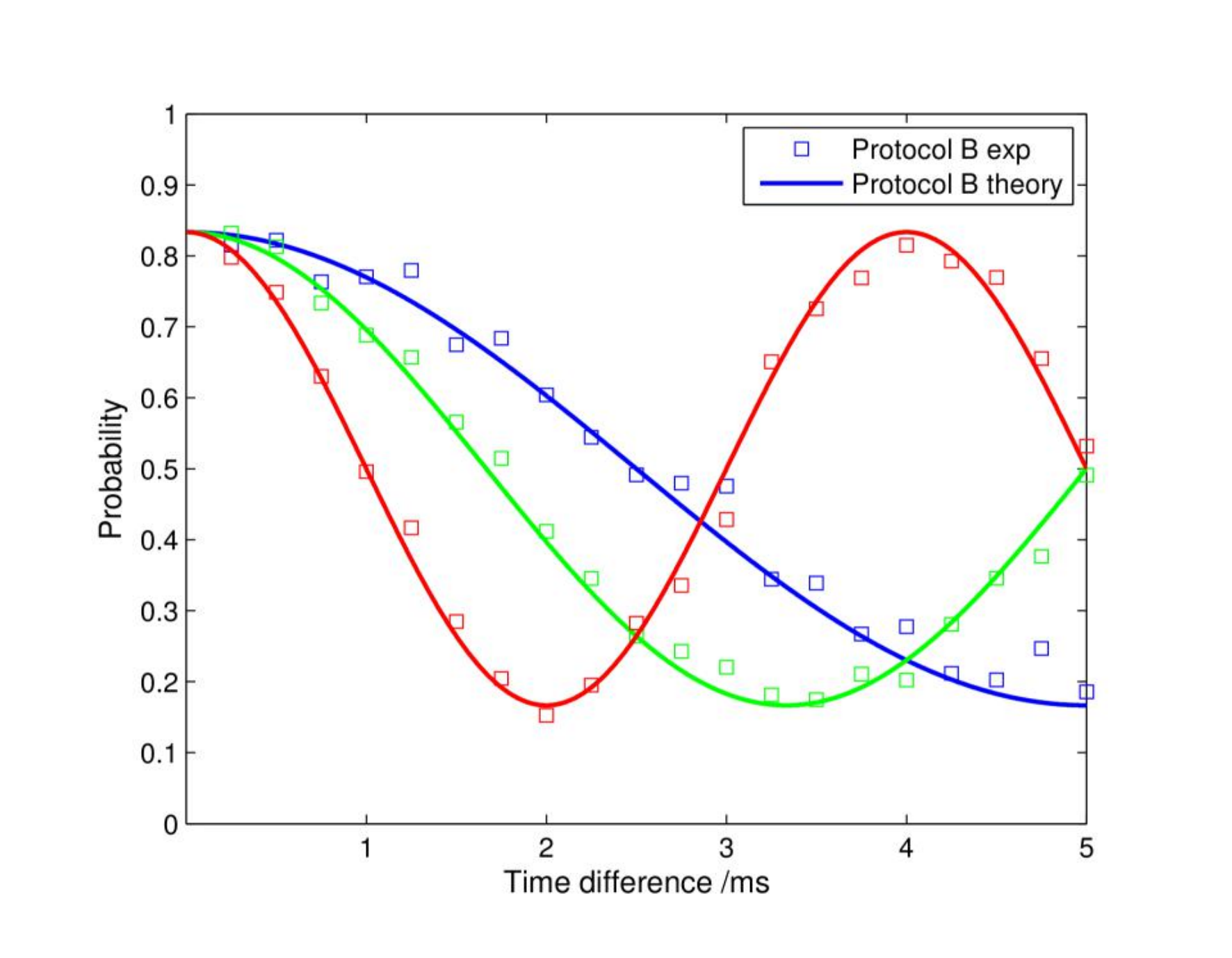}
\end{center}
\setlength{\abovecaptionskip}{-0.00cm}
\caption{\footnotesize{\textbf{Synchronize the other three clocks in one experiment using protocol B.} And the red color stands for Bob with $\omega_2/2\pi$=250Hz,  the green color stands for Charlie with $\omega_3/2\pi$=150Hz and the blue color stands for David with $\omega_4/2\pi$=100Hz. Moreover, the line is the probability value in theory with different $\omega$ showed in Eq. (\ref{eq:4bit-zstate-result}). The red squares are the experimental results of protocol B with 20 time difference $\Delta$ of Bob. Similarly the green squares are the experimental results of Charlie and the blue squares are the experimental results of David.}} \label{fig:result2}
\end{figure}

Compared to the two-parties QCS, the advantages of the multiparty QCS is that it can synchronize the other clocks in one experiment except the standard clock. In order to show that, we have to choose different $\omega$: $\omega_2/2\pi=250Hz,\omega_3/2\pi=150Hz,\omega_4/2\pi=100Hz$.
 We take the protocol B for example. Finally, we will get the probabilities from the corresponding $^{13}$C-labeled, and compare these points with the theory, showed in Fig. \ref{fig:result2}.

\begin{table}[tbp!]
\centering
\caption{\footnotesize{Synchronize Bob's clock using different protocol.}} \label{table:Std deviation}
\begin{tabular}{lccr}
\hline
\hline
$\omega$/2$\pi$&100&150&250\\
\hline
Standard deviation of protocol A/$\mu$s &90&54&42\\
Standard deviation of protocol B/$\mu$s &67&29&32\\
\hline
\hline
\end{tabular}
\end{table}
The three curves in Fig. \ref{fig:result2} are theoretical probabilities which represent Eq. (\ref{eq:4bit-zstate-result}) with selecting the corresponding frequency. The squares of different colors stand for the probabilities of the others in experiment. It is clear that three experimental results are in good agreement with the theory. It proves that we can use the multiparty protocol to coordinate all the clocks in one experiment. Due to Eq. (\ref{eq:4bit-wstate-result}-\ref{eq:4bit-zstate-result}), we can derive the time difference $\Delta'$ from the probabilities, compare them with the real time difference $\Delta$ , and get the standard deviation of the two protocols further showed in Table \ref{table:Std deviation}. We can find that the protocol B has less errors and the higher frequency is , the greater accuracy will be with ignoring experimental errors.

\section{conclusion}
\label{conclusion}
In summary, we overview the different QCS protocols and introduce the two multiparty QCS protocols. We demonstrate the multiparty QCS protocols in four-qubit NMR system. Moreover, the results getting  from our experiments are in good agreement with the theory. It also proves that protocol reported by Ben-Av has a better accuracy than the protocol reported by krco. Our experiment is the first successful attempt which is an excellent demonstration of the feasibility of the multiparty QCS.

However, there also exist some questions. As proved in the theory and demonstrated in the experiment, a higher frequency is good for the synchronization accuracy. But an unitary operation in the NMR system should be realized in several or dozens of microseconds. So if we want to verify more accurate in the multiparty protocol, we should have to choose another quantum system.

In order to verify the two multiparty QCS protocols, we set the same time difference $\Delta$ in one experiment between the standard clock and the others though we have completed the experiment for twenty times with selecting different $\Delta$. As we all know in the NMR system, we can obtain only one free induction decay (FID) signal for one experiment. Moreover we read out the experiment results from the frequency spectrum which is the fourier transform of the FID. Due to the particularity of NMR system, if there are different $\Delta$ between the standard clock with the others, we can repeat the experiment three times with the $\Delta$  in step 3 of Fig. \ref{fig:quantum circuit} assigned with corresponding values.

\begin{acknowledgments}
{\bf Acknowledgments.} We are grateful to
to the following funding sources: National Natural Science
Foundation of China under Grants No. 11175094 and No.
91221205; National Basic Research Program of China under
Grant No. 2015CB921002.
\end{acknowledgments}

\end{document}